\input phyzzx
\vsize 9.25in
\hsize 6.3in
\def\scri{\cal I}
\def\hg{\hat g}
\def\hR{\hat R}
\def\hn{\hat \nabla}
\def\vf{\vec f}
\def\cR{\cal R}
\rightline{UATP-95/03}
\rightline{October 1995}
\vskip 0.2in 
\centerline{\seventeenbf Do Naked Singularities Form?}
\vskip 0.75in
\centerline{\caps Cenalo Vaz\footnote{\dagger}{Internet:
cvaz@mozart.si.ualg.pt}}
\centerline{\it Unidade de Ci\^encias Exactas e Humanas}
\centerline{\it Universidade do Algarve}
\centerline{\it Campus de Gambelas, P-8000 Faro, Portugal}
\vskip 0.2in
\centerline{and}
\vskip 0.2in
\centerline{\caps Louis Witten\footnote{\dagger\dagger}{Internet:
witten@physunc.phy.uc.edu}}
\centerline{\it Department of Physics}
\centerline{\it University of Cincinnati}
\centerline{\it Cincinnati, OH 45221-0011, U.S.A.}
\vskip 0.75in
\centerline{\bf \caps Abstract}
\vskip 0.2in

\noindent A naked singularity is formed by the collapse of a Sine-
Gordon soliton in 1+1 dimensional dilaton gravity with a negative
cosmological constant. We examine the quantum stress tensor
resulting from the formation of the singularity. Consistent
boundary conditions require that the incoming soliton is
accompanied by a flux of incoming radiation across past null
infinity, but neglecting the back reaction of the spacetime leads
to the absurd conclusion that the total energy entering the system
by the time the observer is able to receive information from the
singularity is infinite. We conclude that the back reaction must
prevent the formation of the naked singularity.
\vfill
\eject

Roughly speaking naked singularities are singularities that may be
seen by physically allowed observers. So far they have been
considered intuitively undesirable and this has produced a
conjecture that forbids their existence, the so-called cosmic
censorship conjecture. The conjecture simply states that a
physically reasonable system of energy will not, under reasonable
initial conditions, evolve into a naked singularity. The original
intent of the conjecture was of course to prohibit the formation of
naked singularities in classical general relativity so that, for
example, a non-rotating system would radiate away all multipole
moments sufficiently rapidly that the final state would be a
Schwarzschild black hole. However, there are many counter examples
in the literature where naked singularities do indeed form from
classically reasonable initial conditions.${}^1$

The existence of naked singularities is still difficult to
understand physically, and several suggestions have been made to
maintain the viability of the cosmic censorship hypothesis. One is
merely to reiterate that it holds classically but needs some fine
tuning to describe the set of physically reasonable set of states
from which a collapse may begin. The fine tuning would be such as
to exclude all known examples of the formation of naked
singularities. A second way to preserve the conjecture is to argue
from the point of view of string theory.${}^2$ String theory is
said to solve the short distance problems of general relativity by
providing a fundamental length scale proportional to the inverse
square root of the string tension. The expectation is therefore
that the high energy behavior of string theory (in weak coupling)
would forbid any observations of processes associated with high
field gradients over short distances. Yet a third possibility,
which we have argued for elsewhere,${}^3$ is in contradistinction
to the string argument. It is that when the singularity forms the
accompanying Hawking evaporation is huge and quantum gravity or the
back reaction must prevent the singularity from forming.

In this letter we present a model of collapse leading to the
formation of a naked singularity and show that, at least in the one
loop approximation and without taking the back reaction into
account, the Hawking radiation is incoming and infinite. This
suggests, we argue, that because of quantum gravity effects or the
back reaction the singularity will not form but the energy will
dissipate, perhaps explosively. The model we consider is 1+1
dimensional gravity with Sine-Gordon solitons${}^4$ described by
the action
$$S~~ =~~ {1 \over {2\pi}}~ \int d^2 x \sqrt{-g} \left[ e^{-2\phi}
\left( -R~ +~ 4 (\nabla \phi)^2~ +~ \Lambda\right)~ -~ {1 \over 2}
(\nabla f)^2~ +~ 4 \mu^2 e^{-2\phi} (\cos f~ -~ 1) \right]
\eqno(1)$$
where $\phi$ is the dilaton, $f$ is the matter field and $\Lambda$
is the cosmological constant which we take here to be negative,
$\Lambda = - 4 \lambda^2$. $R$ is the scalar curvature, and
$g_{\mu\nu}$ is the two-metric. Our conventions are those of
Weinberg.${}^5$

It is simplest to analyze the field equations in the conformal
gauge, where the metric has the form
$$g_{\mu\nu}~~ =~~ e^{2\rho} \eta_{\mu\nu}\eqno(2)$$
and in lightcone coordinates, $x^\pm = x^0 \pm x^1$. If we require
that the metric returns to the linear dilaton vacuum in the absence
of the incoming soliton, the solution to the field equations is
given by
$$\eqalign{f_{kink}~~ &=~~ 4 \tan^{-1} e^{(\Delta~ -~ \Delta_0)}\cr
\sigma~~ &=~~ \lambda^2 x^+ x^-~ -~ 2 \ln \cosh(\Delta~ -~
\Delta_0)\cr} \eqno(3)$$
where $\sigma = e^{-2\rho} = e^{-2\phi}$ and $\Delta = \gamma_+ x^+
+ \gamma_- x^-$ with 
$$\gamma_\pm~~ =~~ \pm~ \mu {\sqrt{{1 \pm v} \over {1 \mp v}}}.
\eqno(4)$$
Also, $v$ is the velocity of the soliton, $f(x,t) = f(x+vt)$, and
$\Delta = \Delta_0$ is its center. We take  $\Delta_0 > 0$
throughout, without any loss of generality.

The curvature singularity is at $\sigma = 0$, as can be verified
directly by examining the expression ($R = 4\sigma \partial_+
\partial_- \ln \sigma$) for the curvature scalar. The Kruskal
diagram displayed in figure I shows the singularity along with the
trajectory of the soliton center. In this letter we will be
particularly interested in the lower branch of the singularity. It
is made up of a spacelike piece in the far future joined smoothly
to two timelike singularities on either end, being null in the
neighborhood of $\scri^-$, and intersecting $\scri_R^-$ at $x^- = -
\infty, x^+ = 2 \gamma_-/\lambda^2$ and $\scri_L^-$ at $x^+ = -
\infty, x^- = - 2 \gamma_+/\lambda^2$.  The upper branch cannot be
regarded as a true ``formation'' of a naked singularity as the
soliton energy and singularity appear in the spacetime together,
the soliton emerging from the singularity and travelling all the
way to $i^+$. This case has been examined in ref.[6].

Classical soliton energy entering the spacetime therefore produces
a naked singularity in the future. Accompanying the production of
the singularity is Hawking radiation. The stress tensor of this
radiation is completely determined by the trace anomaly and the
conservation equations. The trace anomaly, in turn, can depend only
on the scalar curvature, $R$, as this is the only curvature
invariant with the correct length dimension. The conservation laws
determine the stress tensor up to two functions, one of $x^+$ and
the other of $x^-$ according to
$$\eqalign{\langle T_{++}\rangle~~ &=~~ T^f_{++}~ +~ T^q_{++}~~ =~~
T^f_{++}~~ -~~ \alpha \left[ {{\partial^2_+ \sigma} \over \sigma}~
-~ {1 \over 2} \left[{{\partial_+ \sigma} \over \sigma}\right]^2
\right]~ + A(x^+) \cr \langle T_{--}\rangle~~ &=~~ T^f_{--}~ +~
T^q_{--}~~ =~~ T^f_{--}~~ -~~ \alpha \left[ {{\partial^2_- \sigma}
\over \sigma}~ -~ {1 \over 2} \left[{{\partial_- \sigma} \over
\sigma}\right]^2 \right]~ +~ B(x^-)\cr \langle T_{+-}\rangle ~~
&=~~ T^f_{+-}~~ +~~ \alpha \partial_+ \partial_- \ln \sigma, \cr}
\eqno(5)$$
where $\alpha$ is a positive, dimensionless (spin dependent)
constant and $T^f_{\mu\nu}$ is the classical soliton stress energy
$$\eqalign{T^f_{++}~~ &=~~ {1 \over 2} (\partial_+ f)^2~~ =~~ {{4
\gamma_+^2} \over {\cosh^2(\Delta - \Delta_0)}}\cr T^f_{--}~~ &=~~
{1 \over 2} (\partial_- f)^2~~ =~~ {{4 \gamma_-^2} \over
{\cosh^2(\Delta - \Delta_0)}}\cr T^f_{+-}~~ &=~~ \mu^2 (\cos f~ -~
1)~~ =~~ -~ {{2 \mu^2} \over {\cosh^2(\Delta - \Delta_0)}}. \cr}
\eqno(6)$$
Appropriate boundary conditions must now be imposed to determine
these arbitrary functions. We expect that the tensor is regular at
all points in the spacetime except at the singularity, $\sigma =
0$. Na\"\i vely it may seem that natural conditions are (a) there
is no incoming flux of energy other than the soliton's and (b) the
Hawking radiation vanishes in the absence of the classical soliton
stress energy, i.e., when $\mu = 0 = \Delta_0$. However, one finds
that the tensor satisfying these conditions is not regular
everywhere. This is because requiring the absence of incoming
energy on $\scri^-$ (other than the soliton, condition (a)) implies
that
$$A(x^+)~~ =~~ -~ {1 \over {2 x^{+'2}}},~~~~~~~~~~ B(x^-)~~ =~~ -~
{1 \over {2 x^{-'2}}}\eqno(7)$$
where
$$\eqalign{x^{+'}~~ &=~~ x^+~ -~ {{2\gamma_-} \over {\lambda^2}}
\cr x^{-'}~~ &=~~ x^-~ +~ {{2 \gamma_+} \over {\lambda^2}}, \cr}
\eqno(8)$$
As the term in square brackets in (5) is finite everywhere, no
incoming flux on $\scri^-$ implies that $\langle T_{++} \rangle
\rightarrow -  \infty$ on the lightlike line $x^{+'} = 0$ and
$\langle T_{--} \rangle \rightarrow - \infty$ on the lightlike line
$x^{-'} = 0$. This behavior is physically unacceptable on the
grounds that there is nothing special about the spacetime along
these lines and infinite negative fluxes to the left and right are
in violation of the positive energy conditions. 

On the other hand the tensor that is regular everywhere within the
spacetime (except at the singularity, $\sigma = 0$) and vanishes in
the absence of the soliton is given by
$$\eqalign{\langle T_{++}\rangle~~ &=~~ T^f_{++}~ +~ T^q_{++}~~ =~~
T^f_{++}~~ -~~ \alpha \left( {{\partial^2_+ \sigma} \over \sigma}~
-~ {1 \over 2} \left[{{\partial_+ \sigma} \over \sigma}\right]^2
\right)~ -~  {\alpha \over {2 x^{+2}}}\cr \langle T_{--}\rangle~~
&=~~ T^f_{--}~ +~ T^q_{--}~~ =~~ T^f_{--}~~ -~~ \alpha \left(
{{\partial^2_- \sigma} \over \sigma}~ -~ {1 \over 2}
\left[{{\partial_- \sigma} \over \sigma}\right]^2 \right)~ -~ 
{\alpha \over {2 x^{-2}}}\cr \langle T_{+-}\rangle ~~ &=~~ T^f_{+-
}~~ +~~ \alpha \partial_+ \partial_- \ln \sigma. \cr} \eqno(9)$$
In coordinates that are manifestly asymptotically flat
$$\eqalign{x^+~~ =~~ -~ {1 \over \lambda} e^{-\lambda \sigma^+}~~
+~~ {{2\gamma_-} \over {\lambda^2}} \cr x^-~~ =~~ -~ {1 \over
\lambda} e^{-\lambda \sigma^-}~~ -~~ {{2\gamma_+} \over
{\lambda^2}}\cr}\eqno(10)$$
the tensor approaches 
$$\eqalign{T^{(\sigma)}_{--}~~ \rightarrow~~ 0,~~~~~~~~~~
T^{(\sigma)}_{+-}~~ \rightarrow~~ 0\cr T^{(\sigma)}_{++}~~
\rightarrow~~ {{\alpha \lambda^2} \over 2} \left[ 1~~ -~~ {1 \over
{\left( 1~ -~ {{2\gamma_-}\over {\lambda}} e^{\lambda \sigma^+}
\right)}} \right]\cr} \eqno(11)$$
on $\scri_R^-$ and 
$$\eqalign{T^{(\sigma)}_{++}~~ \rightarrow~~ 0,~~~~~~~~~~
T^{(\sigma)}_{+-}~~ \rightarrow~~ 0\cr T^{(\sigma)}_{--}~~
\rightarrow~~ {{\alpha \lambda^2} \over 2} \left[ 1~~ -~~ {1 \over
{\left( 1~ +~ {{2\gamma_+}\over {\lambda}} e^{\lambda \sigma^-}
\right)}} \right]\cr} \eqno(12)$$
on $\scri_L^-$. Therefore consistency seems to require an incoming
flux of radiation across past null infinity. This flux is seen to
increase smoothly from zero on $i^-$ to the constant value $\alpha
\lambda^2 /2$ as the lightlike lines $x^- = -2\gamma_+ /\lambda^2$
and $x^+ = 2\gamma_-/\lambda^2$ are approached. The total energy
flowing into the spacetime is the integrated flux from $i^-$ to the
point $x^+ =  2\gamma_-/\lambda^2$ on the right and $x^- = - 2
\gamma_+ /\lambda^2$ on the left. This is obviously infinite, a
result that is nonsense. It is easy to see that this will occur
whenever the singularity is such as to be visible to a timelike
observer.  This is possible in the soliton model because of the
timelike part of the singularity in the neighborhood of $\scri^-$. 
As the total energy entering the system across past null infinity
cannot be greater than the classical soliton energy, the effect of
the incoming Hawking radiation will be to de-localize the soliton. 
On the other hand, the back reaction can be expected to reduce the
spreading of the soliton across $\scri^-$. This would be possible
only if the back reaction works toward eliminating the timelike
piece of the singularity. Indeed we may conclude that quantum-
gravity effects, the back reaction of the spacetime, will not
permit the formation of the classical singularity as described
earlier and that a simultaneous quantization of all fields in the
system is necessary for even a qualitative understanding of the
process. 

The above arguments have been based on what one might physically
expect of the back reaction of the spacetime. To incorporate the
quantum effects to lowest order in studying the back reaction, one
must include the contribution of the conformal anomaly (which
arises because the measure on the space of matter fields is non-
invariant) in the effective action.  An important question is
therefore whether the model being considered here is consistent,
that is, if the equations of motion that have been solved above are
consistent to lowest order. Below we will briefly show, following
arguments by de Alwis${}^7$ and Bilal and Calan${}^8$, that the
Sine-Gordon theory in (1) is actually a special case of such a
consistent theory. Our arguments will closely follow those of
ref.[7]

We can expect that the full quantum action, including matter but
not including ghosts takes the form (in terms of some fiducial
metric $g = e^{2\rho} \hg$)${}^7$
$$S[X,\hg]~~ =~~ - \int {\sqrt {\hg}} \left[ - {1 \over 2}
\hg^{\mu\nu} G_{ab} \nabla_\mu X^a \nabla_\nu X^b~ -~ \hR \Phi~ +~
T(X) \right] \eqno(13)$$
where $T$ is the tachyon potential and $X^a$ is the $N+2$
dimensional vector
$$X^a~~ =~~ \left[ \matrix{\phi \cr \rho \cr f_1 \cr . \cr . \cr .
\cr f_N} \right] \eqno(14)$$
including the $N$ matter fields. The modified CGHS theory we have
considered is the weak coupling limit of (13), i.e., in the limit
$e^{2\phi} << 1$, (13) should look like
$$\eqalign{S[\phi,\rho,f_i]~~ &=~~ \int d^2 \sigma {\sqrt{\hg}}
\left[ e^{-2\phi} \left( 4(\hn \phi)^2~ -~ 4 \hn \phi \cdot \hn
\rho \right)~ -~ \alpha (\hn \rho)^2 \right. \cr &\left. -~\hR
\left( e^{-2\phi}~ -~ \alpha \rho\right)~ +~ \Lambda e^{2(\rho-
\phi)}\right.\cr &\left. -~ {1 \over 2} \sum_i (\hn f_i)^2~ +~ 4
\mu^2 e^{2(\rho-\phi)}U(\vf)\right]\cr} \eqno(15)$$
where, in the case we are studying, $U(\vf)$ is the Sine-Gordon
potential. Comparing (13) and (15), one finds that the low energy
limit is given by
$$\eqalign{G_{ab}~~ &=~~ \left[\matrix{-8e^{-2\phi} & 4e^{-2\phi}
& 0 & . & . & . & 0 \cr 4e^{-2\phi} & 2\alpha & 0 & . & . & . & 0
\cr 0 & 0 & 1 & . & . & . & . \cr 0 & 0 & 0 & 1 & . & . & . \cr .
& . & . & . & . & . & .\cr. & . & . & . & . & . & 1} \right] \cr
\Phi~~ &=~~ +~ \alpha \rho~ -~ e^{-2\phi} \cr T(X)~~ &=~~ -~
\Lambda e^{2(\rho-\phi)}~ - 4 \mu^2 e^{2(\rho-\phi)} U(\vf). \cr}
\eqno(16)$$
Now consider a generalization of the above field space metric, but
with the appropriate limit
$$ds^2~ =~ -8e^{-2\phi} (1 + h_1(\phi)) d\phi^2~ +~ 8e^{-2\phi} (1
+ h_2(\phi)) d\phi d\rho~ +~ 2\alpha (1 + h_3(\phi)) d\rho^2~ +~
\sum_i df_i^2 \eqno(17)$$
where $h_i(\phi)$ are $O(e^{2\phi})$ funcionals of the dilaton,
$\phi$. The metric can be brought to a flat form by the following
field redefinitions:
$$\eqalign{y~~ &=~~ {\sqrt{2\alpha}} \left[ \rho~ -~ {1 \over
\alpha} e^{-2\phi}~ +~ {2 \over \alpha} \int e^{-2\phi} h_2(\phi)
d\phi \right]\cr x~~ &=~~ \int P(\phi) d\phi\cr}\eqno(18)$$
where
$$P(\phi)~~ =~~ e^{-2\phi} \left[8e^{2\phi} (1 + h_1)~ +~ {8 \over
\alpha} (1 + h_2)^2 \right]^{1 \over 2} \eqno(19)$$
which give for the metric in (17)
$$ds^2~~ =~~ -dx^2~~ +~~ dy^2~~ +~~ \sum_i df_i^2. \eqno(20)$$
In the weak field limit, 
$$\eqalign{y~~ &\sim~~ {\sqrt{2\alpha}} \left( \rho~ -~ {1 \over
\alpha} e^{-2\phi}~ +~ . . . \right) \cr x~~ &\sim~~ 2 {\sqrt{2
\over \alpha}} \left( {\alpha \over 2} \phi ~ -~ {1 \over 2} e^{-
2\phi}~ +~ ... \right)\cr} \eqno(21)$$
Let us now consider the beta-function equations for the action in
(13)
$$\eqalign{0~~ =~~ \beta^G_{ab}~~ &=~~ - {\cR}_{ab}~ +~ 2 \nabla_a
\nabla_b \Phi~ -~ \nabla_a T \nabla_b T~ +~ .... \cr 0~~ =~~
\beta^\Phi~~ &=~~ - {\cR}~ +~ 4 G^{ab} \nabla_a \Phi \nabla_b \Phi~ 
-~ 4 \nabla^2 \Phi~ +~ {{N-24} \over 3}~ +~ . . . . \cr 0~~ =~~ 
\beta^T~~ &=~~ -2 \nabla^2 T~ +~ 4 G^{ab} \nabla_a \Phi \nabla_b T~
-~ 4T~ +~ . . . \cr} \eqno(22)$$
It is easy to see, using (20), that the first of the above
equations implies that $\Phi$ is a linear function of the fields,
and to have the appropriate limit it must be uniquely given by
$$\Phi~~ =~~ {\sqrt{\alpha \over 2}} y \eqno(23)$$
The second beta function equation gives trivially $\alpha = (N-
24)/6$, and the third implies that
$$\partial_x^2 T~ -~ \partial_y^2 T~ +~ {\sqrt{2\alpha}} \partial_y
T~ -~ \sum_i \partial_{f_i}^2 T~ -~ 2T~~ =~~ 0 \eqno(24)$$
where we have retained only first order terms in the tachyon
potential. Consider, first, that $T(X)$ depends only on the fields
$(x,y)$. Obviously, a solution that will reduce to the appropriate
one in the limit of weak coupling is
$$T(x,y)~~ =~~ \Gamma e^{{\sqrt{2\over \alpha}} (y-x)}~~ \sim~~
\Gamma e^{2(\rho-\phi)} \eqno(25)$$
where $\Gamma$ is some constant. Including the $f$ fields, if $T$
is of the form
$$T(X)~~ \sim~~ e^{{\sqrt{2\over \alpha}} (y-x)} U(\vf)$$
i.e., if $T(x,y,f_i)$ has the same coupling to the $\rho$ and
$\phi$ fields as the cosmological term, then (24) can be satisfied
only if $U(\vf)$ is harmonic,
$$\sum_i \partial^2_{f_i} U(\vf)~~ =~~ 0 \eqno(26)$$
If one restricts the number, $N$, of fields, $f_i$, to be even, the
potential 
$$U(\vf)~~ =~~ 4 \mu^2 \prod_{i=1}^{N/2} \cos f_i \prod_{j=N/2+1}^N
\cosh f_j \eqno(27)$$
satisfies the condition (26). Now (13) is linear in $T$ (to this
approximation), so
$$T(X)~~ =~~ -~ \Lambda e^{2(\rho-\phi)}~~ -~~ 4 \mu^2 e^{2(\rho -
\phi)} (\prod_{i=1}^{N/2} \cos f_i \prod_{j=N/2+1}^N \cosh f_j~ -~
1) \eqno(28)$$
is a consistent potential. It is also consistent with the equations
of motion derived from (1) as taking $f_i = 0$ for $i \geq 2$ gives
back the soliton model and, for any $f_i$, $f_i = 0$ is a solution
of the classical equations of motion with this potential. 

In conclusion, the following scenario seems to emerge. Ignoring
quantum effects, the observer would enter spacetime at $i^-$
accompanied by highly localized soliton energy and doomed to
finally crash into a singularity. Before doing so he would have to
cross $x^+ = 2\gamma_-/\lambda^2$ or $x^- = -2\gamma_+ /\lambda^2$
or both. Once he has crossed these lines he would be able to
receive information from the singularity and would realize that the
cosmic censorship hypothesis could not hold in his universe. As we
have seen, however, quantum effects will play an important role.
The stress tensor that is regular throughout the spacetime except
at the singularity itself indicates that on entering the spacetime
the observer encounters a flux of incoming energy across $\scri_L^-
$ and $\scri_R^-$ and accompanying the soliton. Neglecting the back
reaction, the flux appears to increase steadily from $i^-$ leading
to an infinite total energy entering the spacetime {\it before} the
singularity becomes actually visible. This is obviously physically
inviable and the back reaction can be expected to step in, stop the
process and prevent the naked singularity from actually forming.
\vskip 0.25in

\noindent{\bf Acknowledgements}

This work was supported in part by  NATO under contract number CRG
920096. C. V. acknowledges the partial support of the {\it Junta
Nacional de Investiga\c{c}\~ao Cient\'\i fica e Tecnol\'ogica}
(JNICT) Portugal under contracts number CERN/S/FAE/128/94 and
number CERN/S/FAE/1037/95, and L.W. acknowledges the partial
support of the U. S. Department of Energy under contract number
DOE-FG02-84ER40153.
\vskip 0.25in

\noindent{\bf References}
{\item{1.}}P. S. Joshi, {\it Global Aspects in Gravitation and
Cosmology}, Clarendon Press, Oxford, 1993. Chapter 6.

{\item{2.}}E. Martinec, Class. Quant. Grav. {\bf 12} (1995) 941

{\item{3.}}C. Vaz and L. Witten, Phys. Letts. {\bf B325} (1994) 27.

{\item{4.}}Hak-Soo Shin and Kwang-Sup Soh, Phys. Rev. {\bf D52}
(1995) 981.

{\item{5.}}S. Weinberg {\it Gravitation and Cosmology}, John Wiley
\& Sons, Inc., N.Y., (1972).

{\item{6.}}C. Vaz and L. Witten, Class. Quant. Grav. {\bf 12}
(1995) 2607.

{\item{7.}}S. P. de Alwis, Phys. Letts. {\bf B289} (1992) 278;
Phys. Letts. {\bf B300} (1993) 330; Phys. rev. {\bf D46} (1992)
5429.

{\item{8.}}A. Bilal and C. Calan, Nucl. Phys. {\bf 394} (1993) 73.
\vskip 0.25in

\noindent{\bf Figure Caption:}

{\item{\bf 1.}}The Kruskal diagram for $\Lambda = - 4\lambda^2 < 0$
and $\Delta_0 > 0$. Regions II \& IV are physical ($\sigma > 0)$.
Two timelike  singularities are joined smoothly at the soliton
center. The figure was drawn for the following parameter values:
$\mu = 1$, $\lambda = 0.25$, $\gamma_+ = \sqrt{3}$, $\gamma_- = -
1/\sqrt{3}$, $\Delta_0 = 5.0$ and $\alpha = 1/24\pi$.

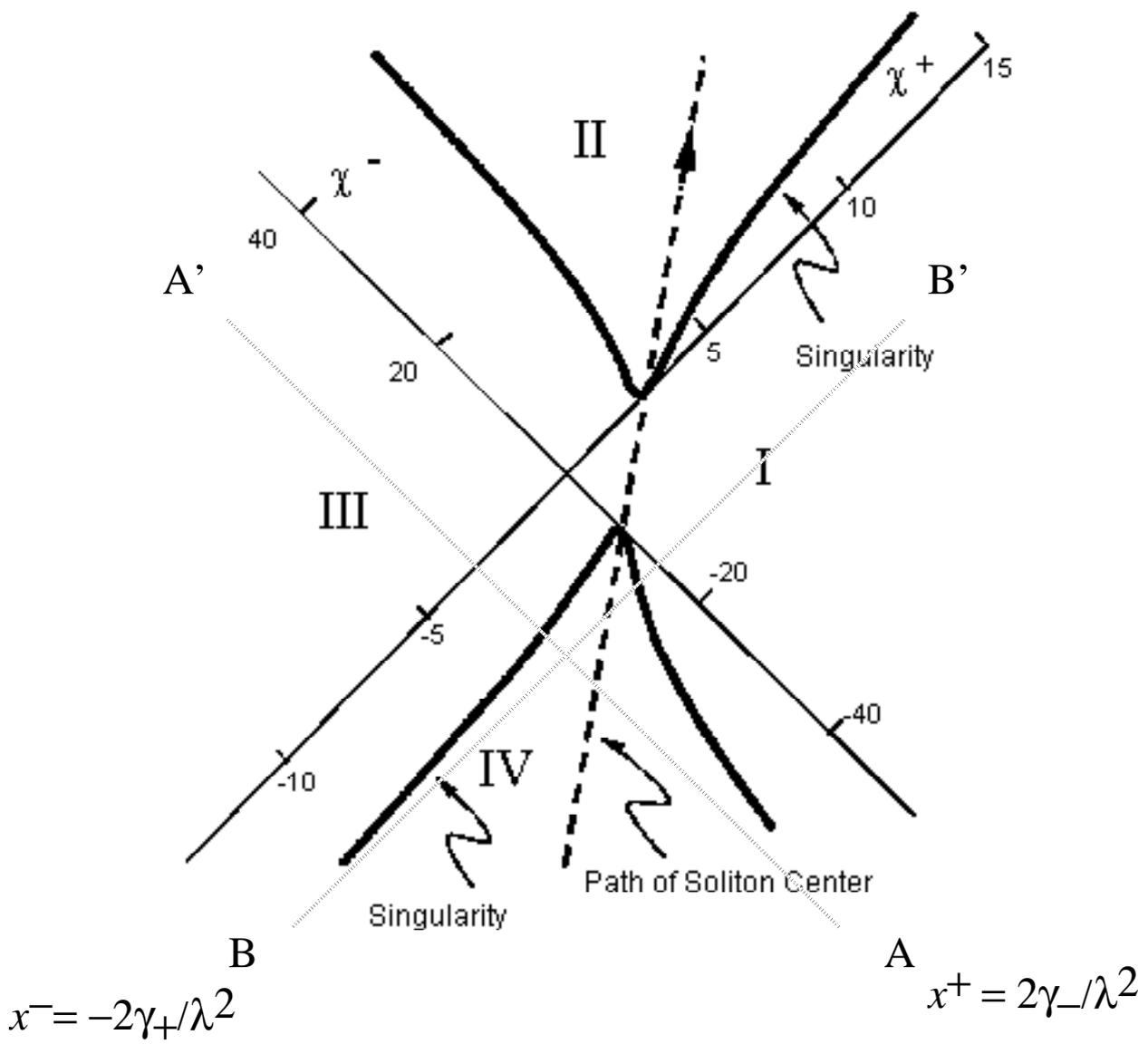
\end